\documentclass[a4paper,10pt]{article}
\usepackage{graphicx}

\title{Subordinated Langevin Equations for Anomalous Diffusion in External
Potentials - Biasing and Decoupled Forces}
\author{S. Eule and R. Friedrich}

\begin{document}

\maketitle

\begin{abstract}
The role of external forces in systems exhibiting anomalous diffusion
is discussed on the basis of the describing Langevin equations. 
Since there exist different possibilities to include the
effect of an external field the concept of {\it biasing} and {\it decoupled}
external fields is introduced. Complementary to the recently established
Langevin equations for anomalous diffusion in a time-dependent external
force-field [{\it Magdziarz et al., Phys. Rev. Lett. {\bf 101}, 210601 (2008)}] 
the Langevin formulation of anomalous diffusion in a decoupled time-dependent 
force-field is derived. 
\end{abstract}
\section{Introduction}
Over the last two decades it has become apparent that many complex
systems exhibit a phenomenom which has been termed anomalous diffusion \cite{Shlesinger, Zaslavsky, MetzlerKlafter}.
On account of this, there has been an increasing interest in stochastic processes
deviating basically from standard diffusion processes characterized by a Gaussian
behavior. Systems exhibiting anomalous diffusion differ from the linear time
dependence of the second moment and rather show $\langle
x^2\rangle\sim t^\alpha$, where $0\leq\alpha\leq 2$. 
In this context processes with $\alpha>1$ dispersing faster than
standard diffusion processes are called superdiffusive while $\alpha<1$ means that 
a system displays subdiffusive behavior.

In the realm of anomalous diffusion, the classical diffusion equation has to be replaced by
the so-called generalized diffusion equations \cite{Balescu}. The most prominent 
representatives of this class of equations are probably the fractional diffusion 
equations \cite{MetzlerKlafter}, where the derivatives with respect to time or
to space or both are replaced by non-integer order derivatives. A more fundamental account
to anomalous diffusion is provided by a stochastic process called Continuous Time
Random Walk (CTRW). This process generalizes the standard Random Walk and allows
for random jump length and random waiting periods between the jumps \cite{Weiss}. It is well-known
that the generalized diffusion equation can be derived from the governing equations
of the CTRW. Another approach to anomalous diffusion has been put forward by
Fogedby who proposed a coupled system of Langevin equations leading to the
generalized diffusion equations \cite{Fogedby}. In a sense, this approach can be considered
as a continuous realization of the CTRW.

In the present paper we consider the effect of external forces onto processes
exhibiting anomalous diffusion. Although the incorporation of external forces
is straightforward in classical diffusion theory, leading to the well-known Fokker-Planck
equations, this task appears to be rather involved when anomalous diffusion is considered.
The arising difficulties are due to the long jumps and the long waiting times that
can occur. Throughout this paper we distinguish between {\it biasing} and 
{\it decoupled} external forces. This notation shall indicate that there are two different
possibilities of the action of the force. When we speak of a biasing field,
we mean that the external field acts as a bias only at the time of the actual
jump. In contrast to this we speak of a decoupled field if the diffusing particle
is affected permanently during the waiting time periods and hence the diffusion process is
decoupled from the effect of the field. Note that this distinction is not necessary
for classical diffusion processes.

While the inclusion of external potentials is relatively well understood on the level
of the generalized diffusion equations and the generalized Fokker-Planck equations
respectively, there are still some open questions as long as the corresponding Langevin
equations are considered. However, an exhaustive comprehension of the Langevin equations
is inevitable to investigate the properties of sample paths of such processes. 

The aim of this paper is to clarify the different possibilities
of including an external force into the framework of anomalous diffusion, namely
the difference between biasing and decoupled external forces, by considering
the corresponding Langevin equations. It is organized as follows. 
First we state some fundamentals concerning the theory of anomalous diffusion and 
thereby shortly review Fogedby's continuous formulation of CTRWs and the concept of subordination.
After shortly reviewing some wellknown and some very recent results on the Langevin formulations 
of the generalized Fokker-Planck equations for biasing external fields 
we establish the Langevin equations for generalized Fokker-Planck equations 
for decoupled external potentials which have not been considered so far. We conclude with a discussion
on the role of external forces in anomalous diffusion.

\section{CTRWs and Generalized Diffusion Equations}

A suited stochastic process to describe discrete sample realizations of many microscopic
processes leading to anomalous diffusion is provided by the Continuous Time Random Walk
(CTRW). This process is an extension of the standard random walk and allows for random waiting times 
between jumps of random length. In the decoupled case the CTRW is characterized by a waiting time distribution
$W(t)$ and a jump length distribution $F(\Delta x)$. Depending on the properties
of these distributions, the CTRW can lead to an anomalous behavior of the 
mean-squared-displacement \cite{Balescu}. The governing equation for the probability
distribution (pdf) of the position of the walker is the Montroll-Shlesinger master equation \cite{MontShle}
\begin{eqnarray}\label{MontShle}
 \frac{\partial}{\partial t}f(x, t)&=&\int \,dx' F(x; x')\int_0^t
 dt'\Phi(t-t') f(x', t') \nonumber \\
& & -\int_0^t dt'\Phi(t-t') f(x, t')\, ,
\end{eqnarray}
where the time kernel $\Phi(t-t')$ is related to the waiting time distribution \cite{phiexpl}. 
The master equation (\ref{MontShle}) has a straightforward interpretation.
It states that the density of particles at position $x$ and time $t$ is
increased by particles that have been at $x'$ at time $t'$ and perform a jump from $x'$ to $x$ at time $t$. 
On the other hand the density is decreased by particles that have been
at $x$ and jump away at time $t$ to some other position. The resulting process is non-Markovian
for waiting time distributions with a power-law tail.

Another account to describe the evolution of pdfs in the context of anomalous
diffusion are the generalized diffusion or fractional diffusion equations.

For a subdiffusive process the generalized diffusion equation can be cast into the form
\begin{equation}\label{genDiffeq}
 \frac{\partial}{\partial t}f(x, t)=\int_0^t dt' \phi(t-t') D
 \frac{\partial^2}{\partial x^2}f(x,t')\, .
\end{equation}
For the special choice $\phi(\tau)=(\tau)^{-1+\alpha}$, which
corresponds to Mittag-Leffler type waiting time distributions $W(\tau)\sim \tau ^{-1-\alpha}$,  
Eq.(\ref{genDiffeq}) yields the (time-) fractional diffusion equation
\begin{equation}
 \frac{\partial}{\partial t}f(x,
 t)=\mathcal{D}_t^{1-\alpha}\frac{\partial^2}{\partial x^2}f(x,t)\, ,
\end{equation}
where $\mathcal {D}_t^{1-\alpha}$ is the Riemann-Liouville fractional derivative. 
It is well-known that generalized diffusion equations can be derived from the 
Montroll-Shlesinger equation, see e.g. \cite{MetzlerBarkai}. 
In order to describe superdiffusive diffusion processes 
the so-called space-fractional diffusion equations have to be taken into account. These
equations describe Markovian processes with power-law distributed jump-length and are often
referred to as L\'evy flights.

\section{Fogedby's Approach and Subordination}

A continuous realization of the  CTRW has been considered by Fogedby in \cite{Fogedby}. His
formulation is based on a system of coupled Langevin equations for the position $x$ and time $t$
\begin{equation}\label{Fogedbysys}
 {\dot x}(s)=\Gamma (s),\qquad {\dot t}(s)=\eta (s)\, ,
\end{equation}
where $\Gamma(s)$ and $\eta(s)$ are random noise sources which are assumed to be independent. 
$\eta(s)$ has in this context to be positive due to causality. The system (\ref{Fogedbysys}) can
be interpreted as a standard Langevin equation in a internal time $s$ that is
subjected to a random time change. This  random time change to the physical 
time $t$ is described by the second equation. The combined process
in physical time is then given according to $x(t)=x[s(t)]$, where $s(t)$ is
the inverse process to $t(s)$ defined as
\begin{equation}
 s({\tilde t})=inf\lbrace s:t(s)>{\tilde t}\rbrace
\end{equation}

Closely related to this concept of Fogedby is the mathematical method of {\it subordination}.
Using the (not to formal) notation of Fogedby one calls the process $x(s)$ a parent process and
$s$ its operational time. The random time-transformation function $t(s)$ has to be a
non-decreasing right-continuous function with an inverse function $s(t)$. The resulting
process in physical time $t$ is then obtained by $x(t)=x[s(t)]$ and is referred to as
subordinated to the parent process. Consequently are the processes $t(s)$ and $s(t)$ named
subordinator and inverse subordinator respectively.

In \cite{Fogedby} it was shown that the Langevin equations (\ref{Fogedbysys}) lead
to a time-fractional diffusion equation if the $\eta (s)$ are governed by a generic one-sided
$\alpha$-stable distribution. Generally the pdf of the subordinated process can be stated
in the form
\begin{equation}\label{subsol}
 f(x,t)=\int_0^\infty ds \, p(s,t)\,f_0(x,s)\, ,
\end{equation}
where $p(s,t)$ is the pdf of the inverse subordinator and $f_0(x,s)$ is the solution of
the parent process \cite{Barkai, Meerschaert}.

\section{Biasing External Force-fields}

Throughout this paper we will restrict to anomalous diffusion processes governed by
waiting time distributions, i.e. we will consider equations of the form (\ref{genDiffeq}).
Hence we consider processes which are e.g. ruled by L\'evy-stable subordinators
and time-fractional equations. The role of external potentials for L\'evy flights is discussed
in \cite{Brockmann}.

Let us first clarify what we mean by {\it biasing} external forces. Therefore, consider 
the generic scenario of a subdiffusive CTRW governed by power-law distributed waiting times.
A biasing external potential or force shall not affect the diffusing particle during the waiting periods but
only provide it with a {\it bias} at the instance of a jump. In a sense one might say that the action of the force
can be regarded as anomalous as well. 

If the considered force is time-independent it is well-known that anomalous diffusion 
in biasing fields can be described by the generalized Fokker-Planck equation
\begin{equation}
 \frac{\partial}{\partial t}f(x,
 t)=\int_0^t dt'\phi(t-t')\left[-\frac{\partial}{\partial x}F(x)
+\frac{\partial^2}{\partial x^2}\right]f(x,t')\, ,
\end{equation}
where $F(x)$ is the external force \cite{MetzlerKlafter, MetzlerBarkai}. The equivalent description based on 
Langevin equations is provided by the coupled system
\begin{equation}
 {\dot x}(s)=F(x)+\Gamma (s),\qquad {\dot t}(s)=\eta (s)\, ,
\end{equation}
where $\Gamma (s)$ is a Gaussian and $\eta (s)$ is a fully skewed
$\alpha$-stable L\'evy noise source \cite{Fogedby}.

\begin{figure}
\begin{center}
\includegraphics[width=0.6\linewidth]{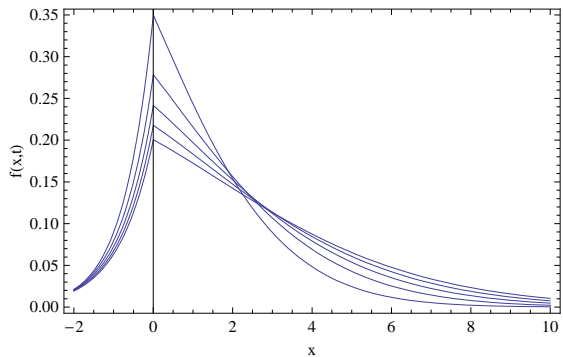}
\caption{Time evolution of the probability density for a constant external biasing force
$F=v$ for Mittag-Leffler type waiting time distributions with $\alpha = 0.5$ for the consecutive
times $t=1$ to 5. One can observe the persistence of the maximum at the origin indicating
that there is no internal dynamic between the waiting times. The external biasing force acts only
at the time of the displacements resulting in a asymmetry of the pdf. The plume stretches more
and more into the direction of the force.}\label{biasadvec}
\end{center}
\end{figure}

If a time-dependent external force $F(t)$  is considered
it turns out that the situation is by far more involved. There exist different alternatives
to include the force. One can for example consider the generalized Fokker-Planck equation
\begin{equation}\label{FFPEwrong}
 \frac{\partial}{\partial t}f(x,
 t)=\int_0^t dt'\phi(t-t')\left[-\frac{\partial}{\partial x}F(t')
+\frac{\partial^2}{\partial x^2}\right]f(x,t')\, .
\end{equation}
However, such a generalized Fokker-Planck equation turns out to be physically meaningless.
The correct equation has been found recently \cite{SokKlaf, Heinsalu, Shushin} 
\begin{equation}\label{timedep}
  \frac{\partial}{\partial t}f(x, t)=\left[-\frac{\partial}{\partial x}F(t)
+\frac{\partial^2}{\partial x^2}\right]\int_0^tdt'\phi(t-t')f(x,t')\, .
\end{equation}
Recall that for the fractional time-kernel $\phi(t-t')=(t-t')^{(1-\alpha)}$, Eq.(\ref{timedep})
yields the fractional Fokker-Planck equation
\begin{equation}
  \frac{\partial}{\partial t}f(x, t)=\left[-\frac{\partial}{\partial x}F(x, t)
+\frac{\partial^2}{\partial x^2}\right]\mathcal{D}_t^{1-\alpha}f(x,t)\, .
\end{equation}
Notice that the difficulty of time-dependent external forces stems from the fact
that in this case the fractional derivative and the Fokker-Planck drift-term
do not commute anymore. On the basis of the generalized Fokker-Planck equation,
the difficulty is due the fact that it is not clear whether the external force
has to depend on $t'$ or $t$. For a detailed treatment of this issue,
we refer the reader to the original papers. At this point, we want to confine ourselves to a 
simple plausibility argument to account for the correct operator ordering which naturally
does not replace a rigorous derivation.

As we have already mentioned, when time-dependent transition amplitudes $F(x; x')$ are considered, 
the question arises whether this amplitude has to depend on $t'$ or $t$ which is equivalent to
operator ordering problem in Eq.(\ref{timedep}). To answer this question, let us consider the
corresponding CTRW governed by the Montroll-Shlesinger equation (\ref{MontShle}).
According to the interpretation of this equation the probability to be at the position $x$
at time $t$ is increased by the particles that jump at time $t$ from some $x'$ to $x$.
Since this jump which occurs at time $t$ is governed by the transition amplitude $F(x;x')$
it is clear that the transition amplitude has to depend on the time of the jump, i.e. $F(x;x',t)$.
Performing the appropriate limit procedure, one obtains Eq.(\ref{timedep}) as
the correct FFPE for time-dependent Fokker-Planck operators.

Consequently, the corresponding Langevin equation for a time-dependent forcing
is not straight-forward to derive.  In fact, it has even been stated in \cite{Heinsalu}
that it is impossible to find a subordination description for time-dependent external
fields. If the force is assumed to depend on the internal time, i.e.
\begin{equation}\label{wronglangevin}
 {\dot x}(s)=F(s)+\Gamma (s),\qquad {\dot t}(s)=\eta (s)\, ,
\end{equation}
which corresponds to a completely subordinated force, the corresponding generalized
Fokker-Planck equation would be Eq.(\ref{FFPEwrong}) and thus Eq.(\ref{wronglangevin}) 
lacks a physical interpretation. 

The appropriate Langevin system has been found recently by Magdziarz and co-workers \cite{Magdziarz}.
They argued that a deterministic force should not be modified by the subordination procedure and 
depend on the physical time $t$ and  proposed the Langevin equations
\begin{equation}\label{klaftersys}
 {\dot x}(s)=F(t(s))+\Gamma (s),\qquad {\dot t}(s)=\eta (s)\, .
\end{equation}
One recognizes that the force depends on the subordination process. Subordination of the process
$x(s)$ then yields for the force term $F[t(s[t])]=F(t)$ since $t(s[t])=t$ and hence the desired dependence
on the physical time. It has been proven in \cite{Magdziarz} that the Langevin equations (\ref{klaftersys}) yields the
same probability distributions as Eq.(\ref{timedep}) and hence that they describe the same process.

\section{Decoupled External Force-fields}

If a particle is assumed to be affected by an external potential throughout
the whole waiting time period and the anomalous diffusion process is independent
of this potential, we speak of a decoupled potential.

It is instructive to consider a simple example where a particle
which is advected by a constant force during the waiting periods and performs jumps
after the waiting periods. 
The pdf of such a process has been proven to be governed by
\begin{equation}\label{advec}
 \left[\frac{\partial}{\partial t}+v\frac{\partial}{\partial x}\right]f(x, t)=
\int_0^t dt' \phi(t-t') \frac{\partial^2}{\partial x^2}f(x-v(t-t'), t')\, ,
\end{equation}
which can be considered as a generalized advection-diffusion equation, where
the advection is normal while the diffusion is anomalous \cite{Eule}. Observe
the retardation of the pdf on the right-hand side which renders the equation
non-local in space. A solution of this equation can be found after passing into 
a co-moving reference frame. The ansatz $f(x, t)=F(\xi, t)$ with the shifted variable $\xi = x- vt$
yields a  generalized diffusion equation for $\xi$
\begin{equation}\label{gendiffxi}
\frac{\partial}{\partial t}F(\xi, t)=\int_0^t dt' \phi(t-t') \frac{\partial^2}{\partial \xi ^2}F(\xi, t')\, ,
\end{equation}
whose solution is given by (see Eq.(\ref{subsol}))
\begin{equation}\label{gendiffsol}
 F(\xi ,t)=\int_0^\infty ds \, p(s,t)\,F_0(\xi ,s)\, ,
\end{equation}
where $F_0$ is the solution of the standard diffusion equation.

\begin{figure}
\begin{center}
\includegraphics[width=0.6\linewidth]{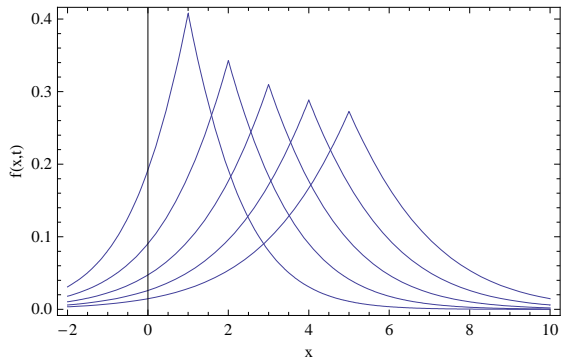}
\caption{Time evolution of the probability density for a decoupled external constant force
with the same settings as in Figure \ref{biasadvec}, i.e. Eq.(\ref{advec}) with a fractional
time-kernel. One can see that this evolution can be regarded as a force-free anomalous diffusion
in a co-moving reference frame. Note that the force does not affect the diffusion process
and hence can be considered as decoupled.}\label{0Umsd}
\end{center}
\end{figure}

In order to establish the corresponding set of Langevin equations, we have to be aware
of the decoupled character of the advective field. That means, the advection
has to be completely independent of the internal time $s$. Let us consider the 
following set of Langevin equations
\begin{equation}\label{adveclangevin}
 {\dot x}(s)=v\,\eta(s)+\Gamma (s),\qquad {\dot t}(s)=\eta (s)\, .
\end{equation}
The solution of the subordinated process $x[s(t)]$ can be found by integration
\begin{equation}
 x(t)=x[s(t)]=\int_0^{s(t)}v\, \eta(s') d[s'(t')]+B[s(t)]\, ,
\end{equation}
where $B[s(t)]$ means subordinated Brownian motion, that is the force-free pure subdiffusive
part of the process \cite{Meerschaert, Magdziarz, Gorenflo, Pira}. The integral can be rewritten as
\begin{eqnarray}
\int_0^{s(t)}v\, \eta(s') d[s'(t')]&=&\int_0^{s(t)}v\, \frac{dt'}{ds'} d[s'(t')] \nonumber \\
&=&\int_0^{t}v\,dt' = v t  \nonumber\\ 
\end{eqnarray}
yielding for the subordinated process
\begin{equation}\label{decoupledsolution}
 x(t)= v t+ B[s(t)] \, .
\end{equation}
Introducing the variable $\xi=x-vt$ again, this equation can be written as
\begin{equation}
 \xi(t)=B[s(t)] \, .
\end{equation}
Thus the variable $\xi$ performs a force-free subdiffusive process and therefore yields the probability
distributions given by Eq.(\ref{gendiffsol}), which proves that the Langevin equations (\ref{adveclangevin})
actually corresponds to the generalized Fokker-Planck equation (\ref{advec}).

The case of time-dependent external field is only slightly more difficult.  Consider
a process, where the particle (of unit mass) performs during the waiting periods an overdamped 
motion according to the equation of motion 
\begin{equation}\label{eom}
 {\dot x}(t)= F(t) 
\end{equation}
where $F(t)$ is some time-dependent force-field.
The corresponding generalized Fokker-Planck equation reads \cite{Eule}
\begin{equation}\label{advectime}
 \left[\frac{\partial}{\partial t}+F(t)\frac{\partial}{\partial x}\right]f(x, t)=
\int_0^t dt' \phi(t-t') \frac{\partial^2}{\partial x^2} e^{-\int_{t'}^t F(t'') dt'' \frac{\partial}{\partial x}}f(x, t')\, .
\end{equation}
The exponential function on the right-hand-side is the so-called Frobenius-Perron operator of the equation of motion
for the deterministic part of $x(t)$. This operator ensures the proper retardation of the probability
distribution during the waiting period \cite{Gaspard}.

Since Eq.(\ref{eom}) describes an invertible conservative system Eq.(\ref{advectime}) can be expressed as (see \cite{Gaspard})
\begin{equation}\label{advectimeretard}
 \left[\frac{\partial}{\partial t}+F(t)\frac{\partial}{\partial x}\right]f(x, t)=
\int_0^t dt' \phi(t-t') \frac{\partial^2}{\partial x^2}f(x-\int_{t'}^tF(t'')dt'', t')\, .
\end{equation}
Performing the ansatz $f(x, t)=F(\xi, t)$ with $\xi=x-\int^t F(t') dt'$, the pdf of $\xi$ is governed
by the generalized diffusion equation (\ref{gendiffxi}).

The corresponding Langevin equation reads
\begin{equation}\label{timedeplangevin}
 {\dot x}(s)=F(s)\,\eta(s)+\Gamma (s),\qquad {\dot t}(s)=\eta (s)\, .
\end{equation}
Integration of this equation yields for the subordinated process
\begin{eqnarray}
 x(t)=x[s(t)]&=&\int_0^{s(t)}F(s')\, \eta(s') d[s'(t')]+B[s(t)] \nonumber \\
& = & \int_0^t F(t') dt' + B[s(t)] \, .
\end{eqnarray}
Evidently $\xi=x-\int^t F(t') dt'$  performs for this case a force-free subdiffusive
process which proves that $x(t)$ is a solution of Eq.(\ref{advectimeretard}).

Note, however, at this point, that the inclusion of space-dependent forces is
only straightforward as long as conservative dynamics is considered because
only in that case the Frobenius-Perron operator can be expressed by a substitution operator
like in Eq.(\ref{advectimeretard}). Even the simple case of a linearly damped motion
between the random kicks, i.e. ${\dot x}=-\gamma x$ leads to a generalized Fokker-Planck
equation whose solution cannot be expressed in a closed form \cite{Eule}. Hence the prove used here
is not applicable anymore. Similarly, a closed form solution of the Langevin equation
cannot be stated for this case.

Comparing the Langevin equation for a biasing time-dependent force Eq.(\ref{klaftersys}) and the 
Langevin equation for the decoupled case, one realizes the difference between these equations.
For the case of a biasing force, the force has to depend on the subordination process in the parent
process. Then the force term yields the contribution $\int_0^t F(t')ds(t')$ to the process. Observe
that the force depends indeed on the physical time $t$ but is integrated over the subordinated measure.
In the decoupled case however, the force is integrated in physical time and thus is completely independent
of the diffusion process. 

\begin{figure}
\begin{center}
\includegraphics[width=0.6\linewidth]{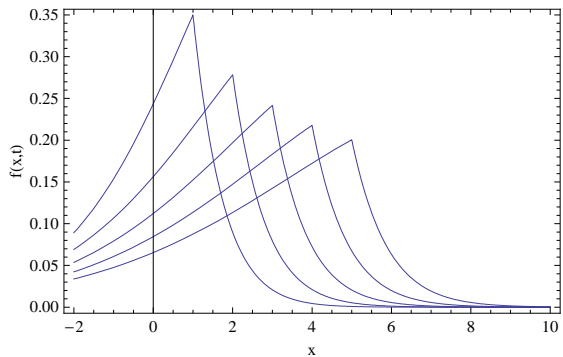}
\caption{Time evolution of the probability density for a decoupled external constant force
combined with biasing constant force of same amplitude but opposite sign. Observe
that the maximum of the distribution moves with the decoupled field while the
distribution becomes more and more asymmetric due to the biasing field.}\label{advecfokker}
\end{center}
\end{figure}

Of course it is possible to state the Langevin equation for a process where a biasing
and decoupled force are acting independently. If $F_B$ denotes the biasing and $F_D$ the decoupled
force the corresponding Langevin equation reads
\begin{equation}\label{langevinboth}
 {\dot x}(s)=F_D(s)\,\eta(s)+ F_B(t(s))+\Gamma (s),\qquad {\dot t}(s)=\eta (s)\, .
\end{equation}
The time evolution of the pdf of such a process is displayed in Fig.
\ref{advecfokker} for a constant biasing force
and a constant decoupled force with same amplitude but opposite sign. Note, however, that in many
settings the two contributions are not independent and thus can display dependences.

\section{Conclusions}
Concluding, in this paper we have discussed the effect of external forces 
on anomalous diffusion processes on basis of their corresponding Langevin
equations. We have introduced the concept of a biasing and a decoupled external
field which has no classical analoge. Corresponding to the recently established
Langevin formulation of biased diffusion in a time-dependent external field \cite{Magdziarz},
we have rigorously derived the Langevin equations for decoupled forces.
To clarify the concept of biasing an decoupled external force in systems
exhibiting anomalous diffusion we have presented the time evolution of
probability densities for the different considered cases. We have shown
that the established Langevin equation for decoupled force fields can be solved exactly
for conservative space-independent dynamics.

The presented work has aimed at a clarification of the role of external forces in complex
systems which are characterized by subdiffusion and long waiting times respectively.
The approach based on the Langevin equation has provided thereby deep insight into the 
physical nature of the processes.

Concluding we shall exemplify the concept by two simple applications each with a constant force. 
First consider the diffusion of tracer particles in an advective flow which 
has frequent obstacles such as e.g. sediments. In this case the external force, i.e.
the advective flow, is decoupled from the diffusion process. Second, if the diffusion of 
groundwater through porous media is examined the gravity field provides a bias on the
 anomalous diffusion process.


\begin{thebibliography}{}
\bibitem{Shlesinger} M. F. Shlesinger, G. M. Zaslavsky and J. Klafter,  Nature (London) {\bf 363}, 31 (1993).
\bibitem{Zaslavsky} G. M. Zaslavsky, Phys. Today {\bf 52}, 39 (1999).
\bibitem{MetzlerKlafter} R. Metzler and J. Klafter, Phys. Rep. {\bf 339}, 1 (2000).
\bibitem{Balescu} R. Balescu, {\it  Aspects of Anomalous Transport in Plasmas}, IOP, Bristol (2005).
\bibitem{Weiss} G. H. Weiss, Aspects and Applications of the Random Walk, Elsevier, Amsterdam (1994).
\bibitem{Fogedby} H. C. Fogedby, Phys. Rev. E {\bf 50}, 1657 (1994).
\bibitem{MontShle} E. W. Montroll and M. F. Shlesinger, in {\it Studies in Statistical Mechanics}, edited
by J. L. Lebowitz and E. W. Montroll (North-Holland, Amsterdam, 1984), Vol. 11.
\bibitem{phiexpl} In Laplace space the time evolution kernel is related to the waiting time by $\Phi(\lambda)=\frac{\lambda W(\lambda)}{1-W(\lambda)}$.
\bibitem{MetzlerBarkai} R. Metzler, E. Barkai and J. Klafter, Europhys. Lett. {\bf 46}, 431 (1999).
\bibitem{Barkai} E. Barkai, Phys. Rev. E {\bf 63}, 046118 (2001).
\bibitem{Meerschaert} M. M. Meerschaert, D. A. Bentson, H.-P. Scheffler and B. Baeumer, Phys. Rev. E {\bf 65}, 041103 (2002).
\bibitem{Brockmann} D. Brockmann, T. Geisel, Phys. Rev. Lett. {\bf 90}, 170601 (2003).
\bibitem{SokKlaf} I. M. Sokolov, J. Klafter, Phys. Rev. Lett. {\bf 97}, 140602 (2006).
\bibitem{Heinsalu} E. Heinsalu, M. Patriarca, I. Goychuk and P. H{\"a}nggi, Phys. Rev. Lett. {\bf 99}, 120602 (2007).
\bibitem{Shushin} A. I. Shushin, Phys. Rev. E {\bf 78}, 051121 (2008).
\bibitem{Magdziarz} M. Magdziarz, A. Weron and J. Klafter, Phys. Rev. Lett. {\bf 101}, 210601 (2008).
\bibitem{Eule} S. Eule, R. Friedrich, F. Jenko and I. M. Sokolov, Phys. Rev. E {\bf 78}, 060102(R) (2008).
\bibitem{Gorenflo} R. Gorenflo, F. Mainardi and A. Vivoli, Chaos, Solitons and Fractals {\bf 34}, 87 (2007).
\bibitem{Pira} A. Piryatinska, A. I. Saichev and W. A. Woyczynski, Physica (Amsterdam) {\bf 349 A}, 375 (2005).
\bibitem{Gaspard} P. Gaspard, {\it Chaos, Scattering and Statistical Mecahnics}, Cambridge University Press, Cambridge (1998).
\end{thebibliography}
\end{document}